\documentclass[useAMS,usenatbib]{mn2e}
\usepackage{epsfig,amssymb}
\title[New physics and astrometry]{
Constraining new fundamental physics with multiwavelength astrometry}
\author[P.~Egorov, M.~Guzinin, H.~Hakopyan and S.~Troitsky]{%
Pavel Egorov$^{1}$, Maxim Guzinin$^{2}$, Hayk Hakopyan$^{2}$ and
Sergey Troitsky$^{3}$\thanks{E-mail: st@ms2.inr.ac.ru}\\
$^{1}$Physics Department, M.V.~Lomonosov Moscow State University,
Vorobjevy Gory, 119991, Moscow, Russia\\
$^{2}$Moscow Institute for Physics and Technology,
Institutskii per.\ 9, 141700, Dolgoprudny, Moscow Region, Russia
\\
$^{3}$Institute for Nuclear Research of the Russian Academy of
Sciences,
60th October Anniversary prospect 7a, 117312, Moscow, Russia
}
\begin{document}
\date{2013 October 16}
\pagerange{\pageref{firstpage}--\pageref{lastpage}} \pubyear{2013}
\maketitle
\label{firstpage}
\begin{abstract}
While the deflection of light is achromatic in General Relativity, it is
not always so in several new-physics models
(e.g.\ certain quantum-gravity and
string-inspired models, models with nonminimal photon-gravity coupling or
with massive photon etc.). We discuss how parameters of these models may be
constrained by precise astrometry at different wavelenghts. From
published observations of the gravitational lens MG~J2016$+$112, we obtain
world-best limits on chromatic gravitational deflection of light
(and the unique limit on the photon mass relevant for distance scales
$>$Mpc). We also outline prospects for further improvement of these
limits.
\end{abstract}


\section{Introduction}
\label{sec:intro}
Though the conventional model of fundamental
physics (which includes the Standard Model (SM) of particle physics to
describe electroweak and strong interactions and General Relativity
(GR) to describe the gravitational one) performs well in explaining
observed phenomena and, up to date, passes most of its experimental tests,
there exist clear indications to its incompleteness (for a review, see
e.g.\ \citet{ST-UFN}), coming both from laboratory experiments (neutrino
oscillations) and from astrophysical observations (dark matter,
accelerated expansion of the Universe and baryon asymmetry). Numerous
theoretical extensions of the standard picture have been suggested which
attempt to solve these experimental problems and/or to reduce fine tuning
of SM parameters. None of the solutions is presently singled out.

Up to now, GR has succefully passed numerous experimental tests (see e.g.\
\citet{Turyshev} for a review). Measurements of the gravitational
deflection of light, performed with high accuracy for astronomical objects
visible close to the Sun, is one of these nice tests, an ``old and good''
one. Gravitational lensing of distant objects is not only a
well-established phenomenon but an important practical tool of
astrophysics and cosmology, see e.g.\ \citet{Lens-Rev}. However, yet
unobserved, modifications of GR have been theoretically proposed in
various contexts. In this study, we attempt to constrain a particular,
though wide, class of the proposed extensions of the ``SM$+$GR''
fundamental model.

The models we focus on predict frequency dependence of the paths followed
by photons in the gravitational field, the so-called ``gravitational
rainbow''. They include, in particular, the following classes of models.
\begin{itemize}
\item
{\it Modifications of GR.} In general, frequency-dependent
corrections arise in any model of quantum gravity, but they are expected
to be suppressed by powers of $(\omega/M_{\rm Pl})$, where $\omega$ is the
photon frequency and $M_{\rm Pl}$ is the fundamental gravity scale (the
Planck mass). For the conventional value of $M_{\rm Pl} \sim 10^{19}$~GeV,
they are hardly observable even in the most precise measurements. However,
there are numerous models on the market where the gravity scale is many
orders of magnitude lower due to the presence of additional space
dimensions (for a review, see e.g.\ \citet{Ru-UFN}). The frequency
dependence of photon paths may arise in certain models inspired by string
theory (e.g.\ \citet{Ellis}), extensions of the minimal gravitational
action \citep{Accioly-gravity}, generalizations of the so-called doubly
special relativity \citep{Smolin} or models formulated with Finsler
geometry \citep{Finsler} etc.
\item
{\it Nonminimal coupling of photons to the gravitational field}, see e.g.\
\citet{Lafrance} and references therein.
\item
{\it Models with massive photon.}
While the SM photon is strictly massless and no indication exists that SM
is wrong in this point, a tiny photon mass may consistently appear in
extended theories either via the Brout-Englert-Higgs mechanism \citep{BE,
H}
or via the St\"uckelberg mechanism \citep{Stuck}. Numerous experimental
constraints on the photon mass are discussed e.g.\ by
\citet{Okun, GammaMassRev}.
\end{itemize}

A similar effect may happen in models with axions of quantum
chromodynamics or similar particles \citep{RaffeltStodolsky} though it is
more difficult to constrain because of its magnetic-field dependence.

Previous studies reported scarce limits on the frequency dependence of the
gravitational deflection of light.
Astrometric limits on the photon mass
from the gravitational deflection of quasar radio signals passing close to
the Sun \citep{PhotonMassOld, PhotonMassNew} are quoted by the
Particle Data Group
\citep{PDG}.
These limits are not the strongest ones; however, in view of model
dependence of many of the constraints, see e.g.\
\citet{PhotonMassNew, Dvali}, they are of independent importance.
\citet{Accioly-gravity} reported constraints on the frequency dependence of
the gravitational deflection of light by the Sun in the context of a
particular modified-gravity model. We are not aware of any other published
constraints. In this work, we improve significantly the limits mentioned
above and sketch prospects for their future improvement.

\section{General estimates}
\label{sec:general}
To put very different models in the frameworks of a single approach,
let $\Delta (\omega)$ be the deflection of light
measured at frequency $\omega$. Then a rather general, though
non-universal, parametrization for the deflection in models we study is (in
the particle-physics units where $\hbar=c=1$ which we use throughout the
paper)
\begin{equation}
\Delta (\omega) = \Delta _0 \left(1\pm\left(\frac{\omega}{M}
\right)^\alpha \right),
\label{Eq:parametrization}
\end{equation}
where $\Delta _{0}$ is the deflection predicted by GR,
$\alpha$ is a model-dependent power (in the models of interest, $\alpha=
\pm 1, \pm 2$) and $M$ is a dimensionful scale expressed through
parameters of the model to be constrained.

There are two ways to constrain $M$ for a given $\alpha$. One is to
compare positions of the source both with and without the deflection thus
measuring $\Delta $ explicitly. Clearly, this approach requires either
a moving deflector like the Sun or Jupiter or a moving light source, a
spasecraft or a planet. The GR expectation, $\Delta _{0}$, may be
precisely calculated in this case, so even a single measurement of
$\Delta$ may constrain $M$. All previously published constraints have
been obtained in this way. Being straightforward, this measurement may
however be performed in a limited number of cases because of particular
trajectories of moving masses in the sky. In the case of the Sun,
its own radiation represents a serious background for close separations.

The second option, which is the subject of this study, is to consider
cases when the gravitational deflection of light is known to be present
but the true direction to the source is unknown (only deflected light is
seen). These include observations of light passing by massive objects
which do not move in the sky. The method may be applied to a wide variety
of sources, at the price of uncertainty in determination of $\Delta_{0}$.
It can be compensated, however, by performing several measurements in one
system: for instance, in this way, observations of multiple images in a
gravitational lens allow to reconstruct the mass distribution and,
indirectly, the true position of the source (assuming GR is valid). For our
purposes, we need to perform observations at different frequencies in
order to eliminate $\Delta_{0}$ and to constrain $M$.

Suppose we performed two measurements, $\Delta(\omega_{1})$ and
$\Delta(\omega_{2})$, of the deflection at frequencies $\omega_{1}$
and $\omega_{2}$, respectively; $\omega_{1}<\omega_{2}$ (see
Sec.~\ref{sec:constraints} for explicit examples). Clearly, we seek tiny
effects and $\Delta(\omega_{1}) \approx \Delta(\omega_{2}) \approx
\Delta_{0}$. We define $k$ as
\begin{equation}
\Delta(\omega_{2}) - \Delta(\omega_{1}) =k  \Delta_{0}
\label{Eq:***}
\end{equation}
and constrain $k$ from observations, $k<k_{+}$ and $k>k_{-}$ (one-sided
limits at a certain confidence level; we suppose in what
follows that GR, $k=0$, is not excluded so $k_{-}<0$ and $k_{+}>0$, true
for our examples). In practice, one expects that
$|k_{\pm}|\Delta_{0}\sim\epsilon$, where $\epsilon$ is the upper limit
on the difference whose expected value is of order the angular resolution.
This would be the least model-dependent result; however, to make it more
transparent, we assume the form (\ref{Eq:parametrization}) for the
corrections to be constrained and express the bound in terms of $M$. To
this end, it is convenient to consider separately the models with
$\alpha>0$ (stronger corrections to GR at high frequencies) and $\alpha<0$
(stronger corrections at $\omega\to0$). Following the particle-physics
jargon, we will call the formers ultraviolet (UV) and the latters infrared
(IR) models. An example of an IR model is any theory with massive photon
while typical UV models are inspired by quantum gravity.

For a UV model, one then obtains the bound
\begin{equation}
M>\left|k_{\pm}\right|^{-1/\alpha} \left(\omega_{2}^{\alpha}
-\omega_{1}^{\alpha} \right)^{1/\alpha} ~~~~ {\rm (UV)},
\label{Eq:*}
\end{equation}
while for the IR case, one has
\begin{equation}
M<\left| k_{\mp}\right|^{-1/\alpha} \left(\omega_{1}^{\alpha}
-\omega_{2}^{\alpha} \right)^{1/\alpha} ~~~~ {\rm (IR)}.
\label{Eq:**}
\end{equation}
In Eqs.~(\ref{Eq:*}), (\ref{Eq:**}), the choice of the upper or lower sign
corresponds to that in Eq.~(\ref{Eq:parametrization}).

To obtain order-of-magnitude estimates, consider three energy bands,
radio ($\omega\sim 10^{-6}$~eV), optical ($\omega\sim 1$~eV) and X-ray
($\omega \sim 10^{3}$~eV), and assume the best corresponding astrometric
accuracies of $\epsilon \sim 10^{-5}$, $10^{-2}$  and
$1$ arcsecond, respectively.
We see, from Eqs.~(\ref{Eq:*}), (\ref{Eq:**}), that, in terms of $M$,
better constraints on IR models may be achieved by observations at two
different radio frequencies while for UV models, the best constraints may
be achieved by comparison of radio measurements with either optical or
X-ray ones. In any case, these estimates are indicative and we should
explore various possibilities for particular sources.

\section{Observational constraints}
\label{sec:constraints}
Here, we sketch two possible practical ways to perform the measurements
outlined in the previous section and give, for each of the two, an example
of the corresponding constraints obtained with a single object. A more
detailed observational study of larger samples of sources will be reported
elsewhere.

\subsection{Gravitational lenses}
\label{sec:lenses}
High-precision measurements of gravitationally lensed systems are
performed with the aim to reconstruct the mass distribution in the lens
which in turn may be important for cosmological applications.
Therefore, there is no lack of observational data, and the precision of
measurements at different frequencies is the guiding rule in the data
selection. Positions of lensed images depend on the mass distribution in a
complicated nonlinear way, and one should expect the same for
potential corrections to the GR formula. However, for small corrections,
we assume Eq.~(\ref{Eq:parametrization}) to be valid for the angular
\textit{distance} between images which we denote
$\overline{\Delta}(\omega)$. For instance, a full calculation
of
$\overline{\Delta}(\omega)$
was performed by \citet{lenses-rainbow} in a
particular modified-gravity model with the result reproducing, in the
small-correction limit, Eq.~(\ref{Eq:parametrization}) with $\alpha=+1$.
The method is then to measure
$\overline{\Delta}(\omega_{1,2})$
for a particular gravitational lens, to constrain $k$ in
Eq.~(\ref{Eq:***}) and to use Eqs.~(\ref{Eq:*}), (\ref{Eq:**}) to obtain
limits on $M$.

We note that for a lens mass distribution with a circular symmetry, the
distance between images $\bar{\Delta}(\omega)$ would be fully equivalent
to the deflection angle $\Delta(\omega)$. The assumption of the circular
symmetry does not hold in a general case, nor in a particular case of the
lens we use below for our estimates. A full analysis of a given lens in
the frameworks of a particular model of anomalous deflection is required
to obtain precise constraints on parameters of this model. Here, we choose
to obtain less precise but model-independent constraints on $k$ and,
consequently, on $M$, which would differ from potential results of more
detailed studies by model-dependent coefficients of order one.

For this study, we have selected the gravitationally lensed quasar
MG~J2016$+$112 which was observed at various frequencies from radio to X
rays with high-resolution instruments. To obtain better constraints on $M$
for both UV and IR theories, we use VLBI observations at
$\omega_{1}=1.7$~GHz, $\omega_{2}=5$~GHz \citep{MG-J-radio} and Chandra
observations at $\omega_{3}\approx 1$~keV \citep{MG-J-X}. The system has
three images A, B, C which are further resolved with VLBI.
Based on the lensing models of \citet{MG-J-radio}, the components of the
C region are more likely to be jet components rather than the quasar cores
and hence, their positions may be sensitive to details of multi-epoch and
multi-frequency data. We therefore consider multifrequency positions of
the A and B components only, taking their principal components, A1 and B1,
in radio data.

To consider IR models, we compare
separations between these components
measured at $\omega_{1,2}$ and presented in
Table~\ref{tab:lens}.
\begin{table}
\centering
\begin{tabular}{|l|c|c|c|}
\hline
\hline
frequency & 1.7~GHz & 5~GHz & keV\\
\hline
B1-A1 (B-A),\hfill  RA& $-3.00574(3)$ & $-3.00595(3)$ & $-2.9(2)$\\
\hfill DEC & $-1.50363(3)$ & $-1.50394(3)$ & $-1.2(2)$\\
\hline
\end{tabular}
\caption{\label{tab:lens}
Angular offsets $\overline \Delta$ (in arc seconds) of the components in the
gravitationally lensed system MG~J2016$+$112 at various frequencies
\protect\citep{MG-J-radio, MG-J-X}. Numbers in parentheses give error bars
in the last digit. }
\end{table}
The positions of radio images do not coincide within the error bars. This
is not surprising since positions of quasar cores are known to be
chromatic \citep{Porkas-chromatic}\footnote{They also may change with time
due to jet proper motions.}. We therefore allow for additional systematic
uncertainties
which we estimate as follows. The offset between A and B is a
two-dimensional vector on the celestial sphere. The two components of this
vector, corresponding to right ascension and declination, are two random
variables. The length of the vector, that is the angular separation we
study, is therefore a random variable which follows the $\chi^{2}$
distribution with two degrees of freedom. We assume this distribution and
allow for an additional systematic error to be added in quadrature to the
statistical errors given in Table~\ref{tab:lens}. The requirement that the
best-fit $\chi^{2}$ corresponds to the $p$-value of 50\% fixes the value
of 0.15~mas for this systematic uncertainty.

The same $\chi^{2}$ distribution allows us to derive directly one-side
confidence intervals for $k$, cf.\ Eq.~(\ref{Eq:***}),
\[
k>k_{-}=-1.9\times 10^{-6} ~~ (95\% ~{\rm CL}, ~ \omega_{1}-\omega_{2}),
\]
\[
k<k_{+}=1.9\times 10^{-4} ~~ (95\% ~{\rm CL}, ~ \omega_{1}-\omega_{2})
\]
(we use $\bar\Delta_{0}\approx \bar\Delta(\omega_{1})$ which is sufficient
for our precision).
For $\alpha=-1,-2$, these intervals transform into constraints on $M$, see
Table~\ref{tab:M}.
\begin{table}
\centering
\begin{tabular}{|c|c|c|c|}
\hline
\hline
$\alpha$ & sign &limit on $M$ (95\% CL), & limit on $M$ (95\% CL),\\
      & & grav.\ lens &  Milky Way\\
\hline
$-2$ &$+$& $<1.6 \times 10^{-9}$~eV & $<2.0 \times 10^{-6}$~eV\\
$-2$ &$-$& $<1.6 \times 10^{-8}$~eV & $<2.0 \times 10^{-6}$~eV\\
$-1$ &$+$&  $<3.2 \times 10^{-12}$~eV & $<6.9 \times 10^{-7}$~eV\\
$-1$ &$-$&  $<3.2 \times 10^{-10}$~eV & $<6.9 \times 10^{-7}$~eV\\
$+1$ &$+$&  $>1.9 \times 10^{4}$~eV & $>3.0$~eV\\
$+1$ &$-$&  $>5.2 \times 10^{3}$~eV & $>3.0$~eV\\
$+2$ &$+$&  $>4.4 \times 10^{3}$~eV & $>1.1$~eV\\
$+2$ &$-$&  $>2.3 \times 10^{3}$~eV & $>1.1$~eV\\
\hline
\end{tabular}
\caption{\label{tab:M}
Constraints on the scale parameter $M$ for different values of $\alpha$
and different signs in Eq.~(\ref{Eq:parametrization}).
}
\end{table}

For the photon mass ($m_{\gamma}=M\sqrt{2}$, $\alpha=-2$, plus sign), this
means
\[
m_{\gamma} < 2.3 \times 10^{-9}~{\rm eV} ~~~ {\rm (95\% CL)}.
\]
This limit is better than the one based on the deflection of light by the
Sun \citep{PhotonMassNew} by two orders of magnitude but is weaker than
some other limits \citep{PDG}. However, one should note that this is the
only existing limit on the photon mass obtained at the distance scale
$>$Mpc. This is important in view of possible dependence both of the photon
mass from the place in the Universe (like in ``chameleon'' models, e.g.\
\citet{chameleon}, or in any model with a non-constant profile of the
Higgs field) and of the obtained limits from the underlying mechanism,
e.g.\ \citet{Dvali, PhotonMassNew, GammaMassRev}.

Turning to UV theories, we, in a similar way, compare measurements at
$\omega_{1}=1.7$~GHz and $\omega_{3}$ (X rays, see Table~\ref{tab:lens}
for data).
In X rays, statistical measurement errors are quite large. We determine
our limits on $k$ by the same method. The assumed systematic error is
$0.16''$. We obtain
\[
k>k_{-}=-0.19 ~~ (95\% ~{\rm CL}, ~ \omega_{1}-\omega_{3}),
\]
\[
k<k_{+}=0.052 ~~ (95\% ~{\rm CL}, ~ \omega_{1}-\omega_{3}).
\]
The corresponding limits on $M$ are, again, given in Table~\ref{tab:M}.
These are the first model-independent (and the world-best for particular
models) limits on the gravitational deflection of light reported in the
literature.

One may wonder whether the radio-interferometric data may be used at all
to constrain effects of the unusual dispersion since the procedure of
reconstruction of the source position assumes the usual dispersion
relation for the detected radio waves. To demonstrate that the obtained
constraints are reliable, we note the following.
Firstly, the gravitational lens we consider does form images of the quasar
(they are observed both with and without the interferometric technique).
Secondly, the following three conditions allow one to
relate the correlation function to the intensity coming from a certain
direction: (1)~the source is far enough, (2)~the emission from different
parts of the (extended) source is not coherent and (3)~the Huygens'
principle works. All these conditions are satisfied even for the massive
photon (in all other cases which we study, the unusual dispersion does not
affect the light propagation between the lens and the observer), thus
justifying the use of the method in principle. Finally, though the image
may be misreconstructed in the case of the nonzero photon mass, it is very
unlikely that the shift in the image due to reconstruction, which is
determined by the geometry of the interferometer, would cancel the
anomalous dispersion effect we attempt to constrain, which is gouverned by
the gravitational field of the lensing galaxy. Moreover, in our case this
potential reconstruction effect is simply too small: the change of the
photon dispersion relation from $|{\bf k}|=\omega$ to  $|{\bf
k}|=\sqrt{\omega^{2}-m_{\gamma}^{2}}$ translates into the effective change
of the frequency, $\omega \to \omega \left(1+m_{\gamma}^{2}/(2\omega^{2})
\right)$, in the expression for the field correlation function. For the
values of $m_{\gamma}$ we constrain and the values of $\omega$ we use,
this correction is of order of $3\%$ of the bandwidth (the latter was
equal to 8~MHz in \citet{MG-J-radio}). This justifies the use of the data
in our case.

\subsection{Deflection in the Milky Way}
\label{sec:milky}
Here we discuss another possibility, which at the present precision level
gives less restrictive constraints as compared to the gravitational
lenses, but may win with the next-generation instruments. The matter
distributed in the Milky Way deflects light rays; once the
distribution of the matter is known, $\Delta_{0}$ may be calculated.
Its value depends on the model of the dark-matter distribution; however,
this dependence is not crucial for our purposes, especially if we compare
observations at different frequencies, thus eliminating $\Delta_{0}$
in Eq.~(\ref{Eq:parametrization}) and leaving it only in the r.h.s.\ of
Eq.~(\ref{Eq:***}) where it is multiplied by a tiny coefficient $k$. The
measurement of the l.h.s.\ of Eq.~(\ref{Eq:***}) is provided by
astrometric measurements of the absolute position of a distant object
performed at two frequencies.

To further understand the technique, one should note that
\textit{absolute} multiwavelength astrometry can hardly reach the required
level of precision because of the unknown systematic offset between
observations at different frequencies. In practice, what is measured is
the relative offset of an object under study with respect to some
calibrators. Since the calibrators should be bright, they are chosen
differently at different wavelengths: for instance, the International
Celestial Reference Frame (ICRF; radio) is determined by
radio quasars while the Hipparcos (optical) frame is related to nearby
stars bright in optical. The positions of the quasars used as ICRF
calibrators are therefore not their true positions, but the deflected
ones: on its way from the source, the light is deflected by the
gravitational field of the Milky Way. It is thus hardly possible to detect
any frequency-dependent gravitational deflection by observations of just
distant radio sources since the same dependence is expected for
calibrators as well. Contrary, the optical reference objects are nearby
stars for which we do not expect any significant deflection by the Milky
Way (they are simply too close to us). These considerations suggest that
to search for the gravitational rainbow, one should measure positions of a
source in both reference frames: optical (not deflected) and radio
(deflected). Presently, the best way is to study those Hipparcos stars
which are radio emitters; an example study was performed by
\citet{RadioStars}. The offset between the optical and radio positions of
the ``radio star'' then constrains $\Delta$ for a particular reference
quasar, while measurements at different radio frequencies would give
constraints on the frequency dependence\footnote{A single-frequency
observation, like that of \citet{RadioStars}, may also be used at the
price of increased systematic uncertainty related to the calculation of
$\Delta_{0}$.}. In particular, \citet{RadioStars} measured ICRF positions
of 46 Hipparcos stars bright in radio, with the best precision of order a
few mas for the angular offset. For instance, one of the most precise
measurements was presented for
U~Sge
(Hipparcos number 94910), for which
the ICRF minus Hipparcos offsets
are $5.1 \pm 7.1$~mas in RA, $4.6\pm
7.2$~mas in DEC. Assuming Gaussian errors, this gives $\epsilon\approx
20$~mas at 95\% CL for $\omega_{1}=8.4$~GHz, $\omega_{2}=5.7 \times
10^{5}$~GHz.
Using the Galactic mass profile by \citet{NFW}, we directly calculated the
estimated GR deflection
$\Delta_{0}\approx 80$~mas for this direction
(details of the calculation and the analysis of other stars will be
presented elsewhere). Eqs.~(\ref{Eq:*}),
(\ref{Eq:**}) then result in the bounds on $M$ listed in
Table~\ref{tab:M}. These bounds are weaker as compared to those obtained
from the gravitational lens because of significantly smaller $\Delta_{0}$
and of the lack of X-ray data.

\section{Conclusions and outlook}
\label{sec:concl}
In this note, we suggested two ways to constrain a certain class of models
which result in chromatic gravitational deflection of light. Both methods
are related to astrometry at different frequencies; one exploits precise
measurements in gravitationally lensed systems while the other one deals
with comparative absolute astrometry of defining sources of optical and
radio reference frames. We illustrated both methods with simple examples
and obtained world-best limits on the chromatic deflection, with the
results listed in Table~\ref{tab:M}.

An interesting application of the study is to constrain the photon mass.
Our study of a particular gravitational lens resulted in the limit
$m_{\gamma} < 2.3 \times 10^{-9}$~eV (95\% CL). This is not the best ever
limit; however, it is the only existing photon-mass constraint relevant
for distance scales larger than Galactic.

The limits we derive may be improved either with a statistical analysis of
larger data samples (which will be reported elsewhere) or with more
precise astrometric measurements. Within the gravitational-lens method,
the limits for IR models (including the photon mass) might be improved
with more precise multifrequency VLBI measurements of the separation
between images of quasars in wide lenses while the key point in improving
limits for UV models is in better angular resolution of X-ray imaging.
Alternatively, important progress is expected in near future of astrometry
with the launch of GAIA which would be able to measure image offsets in
optically bright lenses with the precision of $\sim 20~\mu$as,
overshooting the sensitivity of X-ray studies to $M$ by an order of
magnitude, cf.\ Eq.~(\ref{Eq:*}). An even more dramatic increase of
precision is expected for observations of nearby ``radio stars'' in the
GAIA and ICRF frames; this would make the second method competitive with
the gravitational lenses. A full-sky analysis of this kind may reveal
direction-dependent pattern of differences between the two systems related
to deflection of light of ICRF reference quasars by the Milky-Way
gravitational field. In any case, these next-generation tests would
saturate the precision limit determined by systematic uncertainties.

In case these future studies reveal a significant nontrivial
frequency-dependent effect, its interpretation would require a careful
study both of its influence on the measurement technique and of potential
sources of systematics, including chromatic positions of the quasar cores
(e.g.\ \citet{Porkas-chromatic}), proper motions of the images, effects of
standard-physics dispersion (e.g.\ \citet{QG-vs-standard}) etc.

\section*{Acknowledgments}
The authors are indebted to M.~Fairbairn, D.~Gorbunov, T.~Rashba,
M.~Sazhin, S.~Sibiryakov and O.~Verkhodanov for interesting discussions.
This work was supported in part by RFBR (grants 11-02-01528, 12-02-01203,
13-02-01311 (ST), 12-02-31708 (PE, MG and HH) and 13-02-01293 (all the
team)),  the RF President (grant NS-5590.2012.2, ST, and MK-1170.2013.2,
PE) and the RF Ministry of Science and Education (agreements 8525 and
14.B37.21.0457, ST).

\bsp

\label{lastpage}

\end{document}